\documentclass[11pt,twoside]{article}
\usepackage{asp2010}
\usepackage{graphicx}

\resetcounters

\begin{document}

\title{Turbulence Modelling and Stirring Mechanisms in the Cosmological Large-scale Structure}
\author{Luigi~Iapichino,$^1$ Wolfram~Schmidt,$^2$ Jens~C.~Niemeyer,$^2$ and Johannes~Merklein$^3$
\affil{$^1$Zentrum f\"ur Astronomie der Universit\"at Heidelberg, Institut f\"ur Theoretische Astrophysik, Albert-Ueberle-Str.~2, D-69120 Heidelberg, Germany.\\ E-mail: luigi@uni-heidelberg.de \\
$^2$Institut f\"ur Astrophysik, Universit\"at G\"ottingen, Friedrich-Hund-Platz 1, D-37077 G\"ottingen, Germany \\
$^3$Abteilung Bioklimatologie, Universit\"at G\"ottingen, B\"usgenweg 2, D-37077 G\"ottingen, Germany }}

\begin{abstract}
{\it FEARLESS} (Fluid mEchanics with Adaptively Refined Large Eddy
SimulationS) is a numerical scheme for modelling subgrid-scale
turbulence in cosmological adaptive mesh refinement simulations. In
this contribution, the main features of this tool will be outlined. We
discuss the application of this method to cosmological simulations of
the large-scale structure. The simulations show that the production of
turbulence has a different redshift dependence in the intra-cluster
medium and the warm-hot intergalactic medium, caused by the distinct
stirring mechanisms (mergers and shock interactions) acting in
them. Some properties of the non-thermal pressure support in the two
baryon phases are also described.  
\end{abstract}

\section{Introduction}
\label{introduction}

In the framework of the physics of the cosmological large-scale
structure, turbulent gas flows are an important link between the
thermal and merger history of galaxy clusters, on the one side, and the
non-thermal phenomena (cosmic ray acceleration, amplification of
magnetic fields) in the intra-cluster medium (ICM) on the other. 
A central role in
injecting turbulence in the cosmic flow is played by shocks, which
contribute both to energy dissipation and gas stirring
\citep{mrk00,vbg09}, and by hydrodynamical instabilities triggered by
mergers (e.g., \citealt{hcf03,ias08}). 

The evolution of turbulence in the ICM, as a result of the cluster
merger history, has been explored in many works over the last decade
using hydrodynamical simulations
\citep{rs01,dvb05,in08,vbg11,pim11}. In this contribution we want to
highlight the results of a different approach to the study of
turbulence, adopted by \citet{isn11}. Instead of focusing on a single
cluster, that work followed the evolution of turbulence in a large
cosmological box (with side length of $100\ {\rm Mpc}\ h^{-1}$). This
setup allows to study not only the injection of turbulence in the ICM,
but also in the less dense warm-hot intergalactic medium
(WHIM). Furthermore, the simulation code includes a subgrid scale
(SGS) model for unresolved turbulence \citep{snh06}, coupled to the
adaptive mesh refinement (AMR). The resulting tool, called {\it
  FEARLESS} (Fluid mEchanics with Adaptively Refined 
Large Eddy SimulationS), will be briefly outlined in the next 
section.

\section{Numerical methods}
\label{methods}

{\it FEARLESS} consists of the combination of AMR with a SGS model for the unresolved kinetic energy. Details of this numerical tool have been presented elsewhere \citep{snh06,mis09,isn11}; in essence, the discretization onto a grid of the equations of fluid dynamics is equivalent to applying a {\it filter formalism} \citep{g92} to them. Consequently, additional terms appear in the equations, which take into account the dynamics at unresolved length scales.
 For example, the filtered momentum equation of a viscous, compressible, self-gravitating fluid, becomes
\begin{equation}
\frac{\partial}{\partial t} \langle \rho \rangle \hat{v}_i + \frac{\partial}{\partial r_j}\hat{v}_j\langle \rho \rangle \hat{v}_i = -\frac{\partial}{\partial r_i}\langle p \rangle +\frac{\partial}{\partial r_j}\langle \sigma'_{ij} \rangle \\
+\langle \rho \rangle \hat{g}_i-\frac{\partial}{\partial r_j}\hat{\tau}(v_i,v_j)\,\,,
\label{momentum}
\end{equation}
where $\rho$ is the baryon density, $v_i$ are the velocity components, $p$ is the pressure, $g_i$ the gravitational acceleration and $\sigma'_{ij}$ the viscous stress tensor. Given a variable $f$, with $\hat{f}$ we indicate the application of the filter operator to it (see \citealt{snh06} for the a more rigorous treatment). 

The last term on the right-hand side of equation (\ref{momentum}) contains the turbulent stress tensor $\hat{\tau}(v_i,v_j)$, which accounts for the interaction between the resolved flow and the SGS scales. This term can be expressed in analogy with the viscous stress tensor by means of the so-called {\it eddy viscosity closure} (cf.~\citealt{pope00}), although an improved closure has been recently adopted by \citet{sf11}.

The turbulent stress tensor enters also the definition of the specific filtered kinetic energy, as a contribution from unresolved scales:
\begin{equation}
\hat{e}_{\mathrm{kin}} = \frac{1}{2}\hat{v_i}\hat{v_i} +  \frac{1}{2}\hat{\tau}(v_i,v_j)/ \langle \rho \rangle
\label{ekin}
\end{equation}
It is thus natural \citep{g92} to interpret the trace of $\hat{\tau}(v_i,v_j)/ \langle \rho \rangle$ as the square of the SGS turbulence velocity $q$. This leads to the definition of the SGS turbulence energy $e_{\rm t}$:
\begin{equation}
e_{\mathrm{t}}=\frac{1}{2}q^{2}:= \frac{1}{2}\hat{\tau}(v_i,v_i)/\langle \rho \rangle.
\end{equation}
The SGS turbulence energy is governed by an equation of the following form:
\begin{equation}
\frac{\partial}{\partial t}\langle \rho \rangle e_{\mathrm{t}} + \frac{\partial}{\partial r_j} \hat{v}_j \langle \rho \rangle e_{\mathrm{t}}=\ \mathcal{D}+\Sigma+\Gamma- \langle \rho \rangle (\lambda+\epsilon)\,\,,
\label{eq:etsum}
\end{equation}
The quantities on the right-hand side of equation (\ref{eq:etsum}) determine the evolution of $e_{\mathrm{t}}$ and are the turbulent diffusion term $\mathcal{D}$, the turbulent production term $\Sigma$, the pressure dilatation term $\lambda$ and the viscous dissipation term $\epsilon$. Their closures represent the SGS model.

The method is coupled with AMR
in a way that consistently accounts for
cut-off length scales varying in time and space (see \citealt{mis09}
for a detailed description). The resulting tool is particularly
suitable in the study of turbulence in strongly clumped
media. Furthermore, the cell-wise computation of $e_{\mathrm{t}}$
makes the analysis and visualization of turbulence easy and flexible. 

{\it FEARLESS} has been implemented on the grid-based, AMR hybrid ($N$-body plus hydrodynamical) code Enzo \citep{obb05}. In \citet{isn11}, we ran adiabatic simulations in a cosmological box with a side of $100\ {\rm Mpc}\ h^{-1}$, resolved on a root grid of $128^3$ cells and  $128^3$ $N$-body particles. Four additional AMR levels have been allowed, with a refinement criterion based on overdensity, leading to an effective spatial resolution of $48.8\ {\rm kpc}\ h^{-1}$.

\section{Turbulence and non-thermal pressure support}
\label{turbulence}

\begin{figure}[t]
\begin{center}
\includegraphics[width = 250pt]{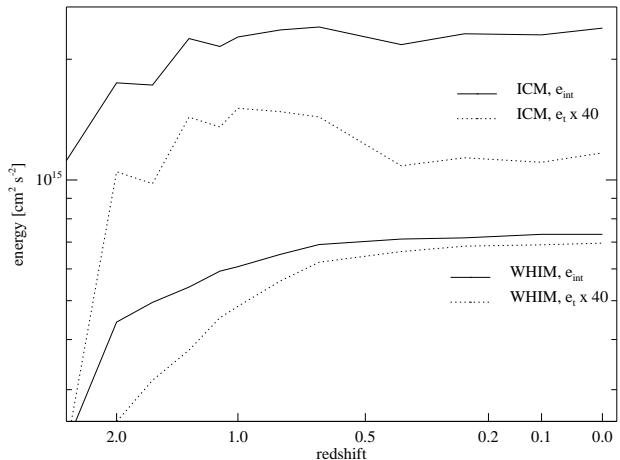}
\end{center}
\caption{Temporal evolution of the mass-weighted averages of the
  internal ($e_{\rm int}$) and turbulence SGS ($e_{\rm t}$) energies
  for the ICM and WHIM baryon phases.
  From \citet{isn11}.}\label{evolution} 
\end{figure}

The definition of the baryon phases in \citet{isn11} is based on a
threshold in temperature $T$ and baryon overdensity
$\delta$. The gas is defined as belonging to the WHIM if $T > 10^5\
{\rm K}$ and $\delta < 10^3$; if $\delta > 10^3$, it belongs to the
ICM. The former phase is found in filaments and cluster outskirts, and
the latter in the denser, collapsed structures. 
In Fig.~\ref{evolution} the time evolution of the average internal and
SGS energy for these two phases is shown. We notice a different
redshift evolution for $e_{\rm t}$: in the ICM, it shows a peak at $z$
between 1.0 and 0.65, followed by a mild decrease. In the WHIM phase,
there is a steady increase to $z = 0$.  

In \citet{isn11} we provide an interpretation of these features in
terms of different stirring mechanisms acting in the two baryon
phases: the evolution of turbulence in the ICM follows closely the
cluster merging history, therefore $e_{\rm t}$ peaks approximately
during the major merger epoch (cf.~\citealt{gms07}) and its decline is
halted by the subsequent minor mergers. Recently, \citet{hj11} studied
the evolution of turbulence in clusters, and found that the fraction
of clusters with large turbulence in the core evolves in time with a
trend very similar to the ICM in Fig.~\ref{evolution}. 
The evolution of turbulence in the WHIM is governed by the gas
accreted on filaments and cluster outskirts, and closely resembles the
evolution of the kinetic energy flux through external shocks
(\citealt{mrk00,soh08}).   

Another interesting problem which was explored in \citet{isn11}
concerns the dynamical pressure support of the cosmic gas. We notice
that, starting from the definition of $e_{\rm t}$, one can identify
the non-thermal pressure caused by unresolved, SGS velocity
fluctuations with $p_{\rm t} = 2/3\ \rho e_{\rm t}$. This term has
been included in an analysis of the dynamical support against
gravitational contraction of the gas  \citep{zff10}. We refer to
\citet{isn11} for a more thorough derivation of the support equations;
here it is sufficient to say that the analysis is based on the
Laplacians of the thermal and dynamical pressure (the latter one
including both resolved and SGS terms).  

It is found that the turbulent support is stronger in the WHIM gas at
baryon overdensities $1 \la \delta \la 100$, and less relevant for the
ICM. A fairly large fraction of the WHIM and ICM gas has a large
vorticity (28.7 and 52.3 per cent in mass at $z = 0$, respectively),
but this is usually associated with an equally large thermal pressure
support, and only in a small volume fraction (of the order of 10
percent) the non-thermal pressure is dynamically significant. 

This result is apparently in contradiction with the idea that the
cluster outskirts, consisting mostly of gas which is newly accreted in
the potential well, could have significant departures from hydrostatic
equilibrium. Our work shows that this is the case only in localized regions, but not globally (see also \citealt{valda11}). More resolved cosmological simulations, focused on single
clusters, will be used to investigate further this point. 

\bibliography{cluster-index}

\end{document}